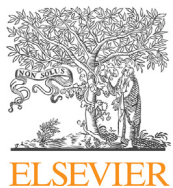
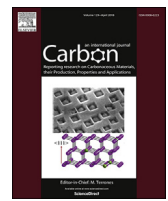

# A hybrid P3HT-Graphene interface for efficient photostimulation of neurons

Mattia L. DiFrancesco, PhD [a, b, 1], Elisabetta Colombo, PhD [a, b, 1], Ermanno D. Papaleo, MD [a, b], José Fernando Maya-Vetencourt, PhD [a], Giovanni Manfredi, PhD [c], Guglielmo Lanzani, PhD [c], Fabio Benfenati, MD [a, b, *]

[a] *Center for Synaptic Neuroscience and Technology, Istituto Italiano di Tecnologia, Largo Rosanna Benzi 10, 16132, Genova, Italy*
[b] *IRCCS Ospedale Policlinico San Martino, Genova, Italy*
[c] *Center for Nano Science and Technology, Istituto Italiano di Tecnologia, Milan, Italy*

## ARTICLE INFO

## ABSTRACT

Graphene conductive properties have been long exploited in the field of organic photovoltaics and optoelectronics by the scientific community worldwide. We engineered and characterized a hybrid biointerface in which graphene is coupled with photosensitive polymers, and tested its ability to elicit light-triggered neural activity modulation in primary neurons and blind retina explants. We designed such a graphene-based device by modifying a photoactive P3HT-based retinal interface, previously reported to rescue light sensitivity in blind rodents, with a CVD graphene layer replacing the conductive PEDOT:PSS layer to enhance charge separation. The new graphene-based device was characterized for its electrochemical features and for the ability to photostimulate primary neurons and blind retina explants, while preserving biocompatibility. Light-triggered responses, recorded by patch-clamp *in vitro* or MEA *ex vivo*, show a stronger light-transduction efficiency when the neurons are interfaced with the graphene-based device with respect to the PEDOT:PSS-based one. The possibility to ameliorate flexible photo-stimulating devices via the insertion of graphene, paves the way for potential biomedical applications of graphene-based neuronal interfaces in the context of retinal implants.

## 1. Introduction

The development of new technologies for the stimulation of neuronal cells and tissues paces with the performance tuning of state-of-art strategies, which flourish thanks to the constant innovation in the fields of nano- or opto-electronics and material science.

Graphene is a 2D material, firstly isolated and identified in 2004 [1], showing outstanding properties, such as one-atom-thickness and consequent low weight/surface ratio, extremely high electrical conductivity, high resistance to stretch and lateral deformation despite a high degree of flexibility, transparency, and compatibility with live cells and tissues. Thanks to these features, graphene has long been studied for applications in the field of biomedical technologies. In the last decades, an increasing interest rose in the scientific community for the development of novel graphene-based scaffolds characterized by biocompatibility [2,3], long-term durability [4], and flexibility [5]. Prototypes under study are wearable devices [6], implantable microelectrodes and microtransistors [7], or *in vivo* cortical, retinal and peripheral nerve probes [8].

On the other hand, graphene has also been extensively exploited to enhance photoconversion processes in organic photovoltaic and photocatalytic devices [9]. In this framework, graphene has been employed as transparent electrode [10], catalytic counter electrode [11], and charge transport layer [12], just to mention a few.

We previously characterized light-sensitive interfaces based on photovoltaic polymers able to modulate the activity of HEK-293 cells [13] and neurons [14,15], which proved to be successful in visual restoration in the Royal College of Surgeons (RCS) rat model of retinal dystrophy [16,17], characterized by photoreceptor degeneration associated with mutation of the *mertk* gene expressed by the retinal pigmented epithelium. This triggers a phagocytosis defect progressive loss of photoreceptors with accumulation of cytotoxic debris in the subretinal space during the first months of postnatal life [18]. For this *in vivo* application, the full-organic implant was composed of a silk fibroin substrate, a poly (3,4-ethylenedioxythiophene)-poly (styrenesulfonate) (PEDOT:PSS) and a photoactive poly-3-hexylthiophene (P3HT) layers, inspired by the core architecture of a typical organic solar cell. In this configuration, P3HT is in direct contact with the electrolytic extracellular fluid, which acts as a cathode together with the tissue itself. Following light-evoked exciton dissociation, the P3HT surface at the tissue interface negatively polarizes and in turn modulates the membrane voltage of neurons grown in contact with the photoactive layer [19].

---

* Corresponding author. Center for Synaptic Neuroscience and Technology, Istituto Italiano di Tecnologia, Italy.
*E-mail address:* fabio.benfenati@iit.it (F. Benfenati).
[1] Equal contribution.

We therefore realized a novel neural interface prototype on a flexible PolyEthylene Terephthalate (PET) substrate by substituting the PEDOT:PSS layer with CVD graphene (G-CVD) in order to exploit the more favourable work function of G-CVD for charge extraction in this architecture, and therefore enhance neuronal photo-modulation. We ran parallel characterizations of the new and the previous prototypes containing PEDOT:PSS from a biophysical point of view, in contact with primary hippocampal neurons and blind retinal explants.

The results exploit the unique performance of graphene for increasing the photoconversion performances of conjugated polymers in a hybrid neural interface that may result in the possibility of reducing the light power threshold needed to trigger neuronal modulation.

## 2. Results

### 2.1. G-CVD enhances charge separation in the P3HT layer under photo-stimulation

Graphene was produced by Chemical Vapour Deposition (CVD) (Graphenea, Spain), transferred to PET substrates (30 μm thickness) and characterized by spectroscopy. Raman spectra show the G and 2D bands typical of a few layers CVD graphene, as highlighted by the two peaks deconvolutions (Fig. 1a). Successively, P3HT (30 mg/ml; Sigma Aldrich) was spin-coated onto the PET/G-CVD structure to a final film thickness of about 100 nm.

The comparison between the optical transmittance of the three-layered devices with G-CVD and PEDOT:PSS as interlayers resulted in very similar spectra, denoting that graphene hydrophobicity did not interfere with the wettability of the substrate before P3HT coating, nor with the P3HT optical properties and absorption range (400–650 nm) (Fig. 1b). Next, we measured the ability of the graphene interlayer to enhance the surface photo-potential by means of a patch-clamp pipette micromanipulated in the close proximity of the device surface in electrophysiological solution. Illumination was provided at 540 nm in order to activate the photosensitive layer P3HT with light close to its absorbance peak [14]. Surface potential recordings in response to green light pulses were realized with the G-CVD and PEDOT:PSS interlayers, both terminated with a P3HT-based heterojunction film in order to emphasize the photo-potential and improve the signal to noise ratio. The recordings showed a significant enhancement of the photogenerated charges in the devices containing graphene with respect to the one with PEDOT:PSS (Fig. 1c), setting the basis for an improved engineering of the previous prototype.

We next performed cyclic-voltammetry measurements in the dark and under illumination at 20 mW/mm$^2$. The neural interface architecture containing G-CVD shows the typical cathodic evolution of a P3HT thin film on a conductive layer (such as on ITO), both in the dark and upon illumination (Fig. 1d). The current density quantification for the two prototypes depicted a strong improvement of the photogenerated current whenever graphene was present as interlayer (Fig. 1e).

We also compared the thermal properties of the two neural interfaces in electrophysiological environment with a patch-clamp pipette and amplifier, exploiting the intrinsic resistance dependence upon temperature (calibrated pipette measurements; inset Fig. 1f) [13]. For both the PEDOT:PSS-based and the G-CVD-based devices, the temperature increase in proximity of the P3HT surface was recorded at increasing light power densities for 50 and 500 ms illumination durations, showing that heating starts to play a relevant role only with long stimuli (500 ms) at high power (>10 mW/mm$^2$). Fully overlapping temperature curves were obtained with the two devices, indicating that light absorption by P3HT is responsible for the light-induced heat generation.

Altogether, these results depict a suitable new architecture of the previous P3HT-based neural interface, with the substitution of the PEDOT:PSS layer with G-CVD able to improve the electrochemical properties of the device upon illumination, without affecting materials integrity or producing excessive overheating.

### 2.2. The G-CVD-based device preserves viability of primary hippocampal cultures

Neuronal adhesion and viability of primary neurons grown in contact with ITO/P3HT:PCBM device was previously reported [14]. We address here whether G-CVD interferes with the properties of the device, by comparing growth and viability of primary hippocampal neurons on PET/G-CVD/P3HT (+G-CVD) devices with those on PET/PEDOT:PSS/P3HT (+PEDOT/PSS) or on a pure PET surface used as control. Hippocampal neurons were cultured for 14 days *in vitro* (DIV) onto the respective substrates coated with poly-L-lysine (PLL), and their viability was estimated as the ratio between fluorescein diacetate-positive and nuclear DAPI labelling-positive cells. As shown in Fig. 2a,b, a high percentage viability resulted from all tested experimental groups (mean ± SEM: PET = 90.62 ± 3.47%, +PEDOT:PSS = 91.58 ± 3.14%, +G-CVD 91.51 ± 1.91%) with no significant changes within the three tested devices (Fig. 2a and b). This indicated that G-CVD did not affect the features of the polymeric device, allowing normal adhesion, growth and survival of neuronal networks.

### 2.3. Light-induced membrane voltage modulation of primary hippocampal neurons is increased in the presence of G-CVD

In view of the increased photo-potential due to the enhanced charge separation in the presence of G-CVD revealed by surface potentials and cyclic voltammetry measurements, we investigated whether the larger light transduction efficiency could be translated into an improved modulation of neuronal activity. Primary hippocampal neurons were cultured *in vitro* on top of either + G-CVD or + PEDOT:PSS devices (in addition to a pure PET control), as described above. Fourteen days after plating (14 DIV), neurons were subjected to patch-clamp measurements in the current-clamp (CC) configuration with two distinct protocols: (i) without current injection (I = 0), allowing the neuronal membrane voltage to set to a spontaneous value (−52,95 ± 1,46 mV PET control, −53,26 ± 1,01 mV + PEDOT:PSS and −52,05 ± 1,19 mV + G-CVD; see Suppl. Figure 1), and (ii) by injecting an outward current to maintain the membrane voltage at −70 mV (I$_{Holding}$ −70 mV) to reproducibly mimic the physiological conditions. Illumination protocols were applied at 1 or 15 mW/mm$^2$ light-power density for 50–500 ms, using a 550/15 nm LED source.

During light stimulation, neurons showed a hyperpolarizing response when measured in the I = 0 modality and a mirror depolarization when using the I$_{Holding}$ −70 mV modality, with no-

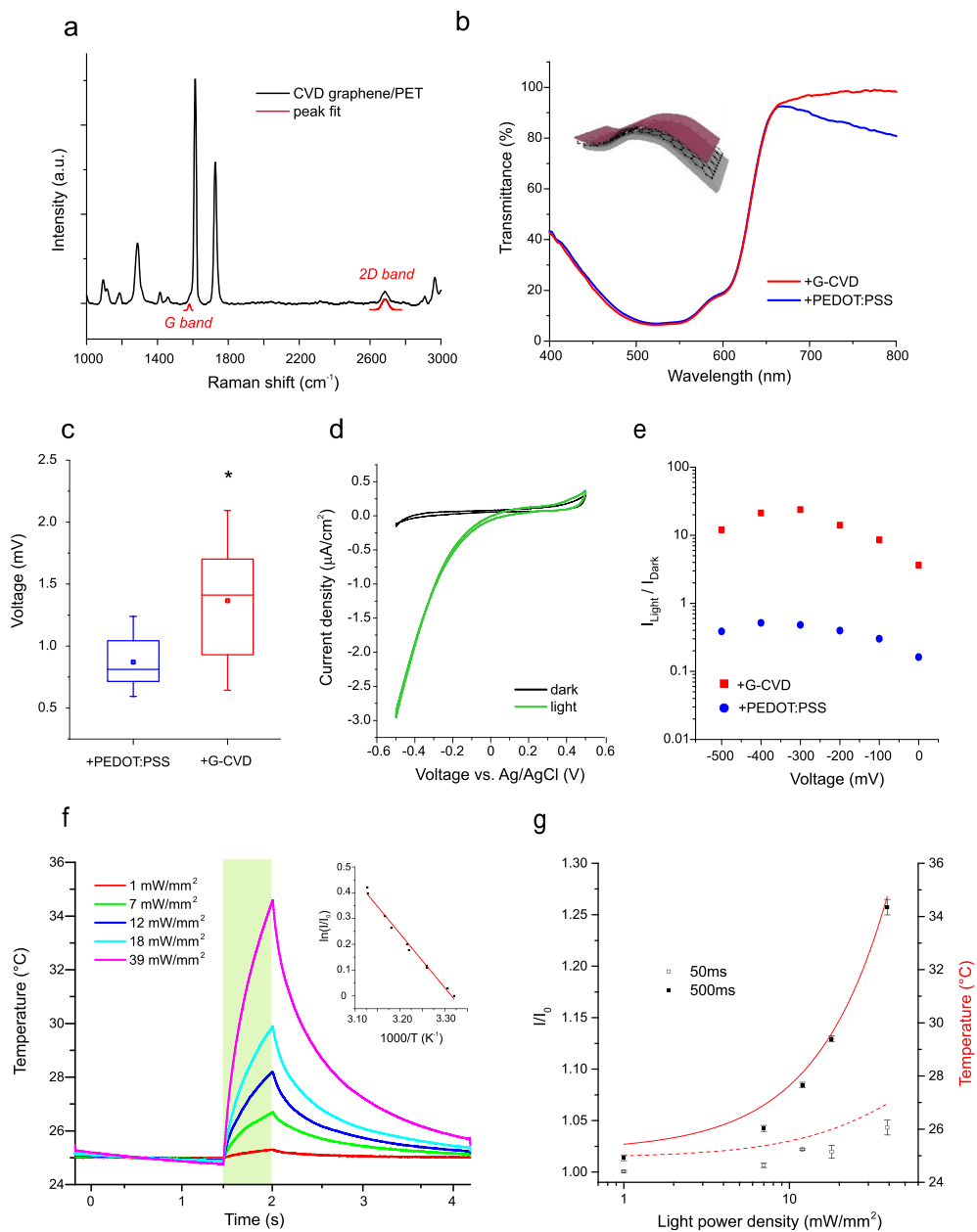

**Fig. 1.** Graphene-based neural interface characterization.
**a**) G-CVD layer transferred onto a PET substrate shows a typical Raman spectrum, where the G and 2D band of graphene clearly emerge from the superimposed PET signal. Both peaks were deconvoluted from the acquired spectrum and their respective Gaussian fit reported. **b**) Optical transmittance of the full device in the visible range shows no interference of the graphene layer with the P3HT main absorption range (400–650 nm). **c**) Comparison of the surface potentials of P3HT:PCBM films on either PEDOT:PSS/PET or G-CVD/PET shows a significantly more pronounced photovoltaic effect on graphene-based devices. The recording is performed with an open patch-clamp pipette (~4 MΩ) in current-clamp at I = 0, micromanipulated in the proximity of the device surface and under illumination at 1 mW/mm² **d**) Cyclic-voltammetry of a P3HT/G-CVD/PET device measured in phosphate-buffered saline (PBS) in the dark and under illumination at 20 mW/mm² and 100 mV/s **e**) The comparison between the cathodic current densities of the two prototypes under study shows a stronger effect on G-CVD-based devices. **f**) Temperature characterization of the P3HT/G-CVD/PET device. The temperature variation in proximity of the device surface under light stimulation of different intensities (540 nm) is monitored by exploiting the intrinsic temperature dependence of a patch-clamp open pipette resistance (calibrated as shown by the Arrhenius plot in the inset). **g**) $I/I_0$ values and the extrapolated temperature variation curve are shown for the experimental conditions used in the following *in vitro* experiments (50 and 500 ms durations). (A colour version of this figure can be viewed online.)

membrane voltage modulation of neurons in contact with the PET control device (Fig. 2c and d). A minority of cells showed an opposite behaviour with depolarization at I = 0 (2.7% of neurons at 15 mW/mm², 500 ms) and hyperpolarization at $I_{Holding}$ −70 mV (14.82% of neurons at 15 mW/mm², 500 ms). Light-induced hyperpolarization and depolarization were measured between the onset and the offset of the light stimulus. To measure net effects of light stimulation, the voltage change measured with neurons on the PET alone control was subtracted from the hyperpolarization and depolarization amplitudes measured with the +G-CVD and the +PEDOT:PSS devices for each stimulation protocol (Δ Device − Δ Ctrl; Fig. 2c and d). The light-dependent membrane voltage modulation resulted significantly enhanced in the presence of G-CVD in most stimulation protocols, both in the I = 0 and the $I_{Holding}$ −70 mV modalities with stronger effects with the latter modality. We concluded that the G-based device modulates

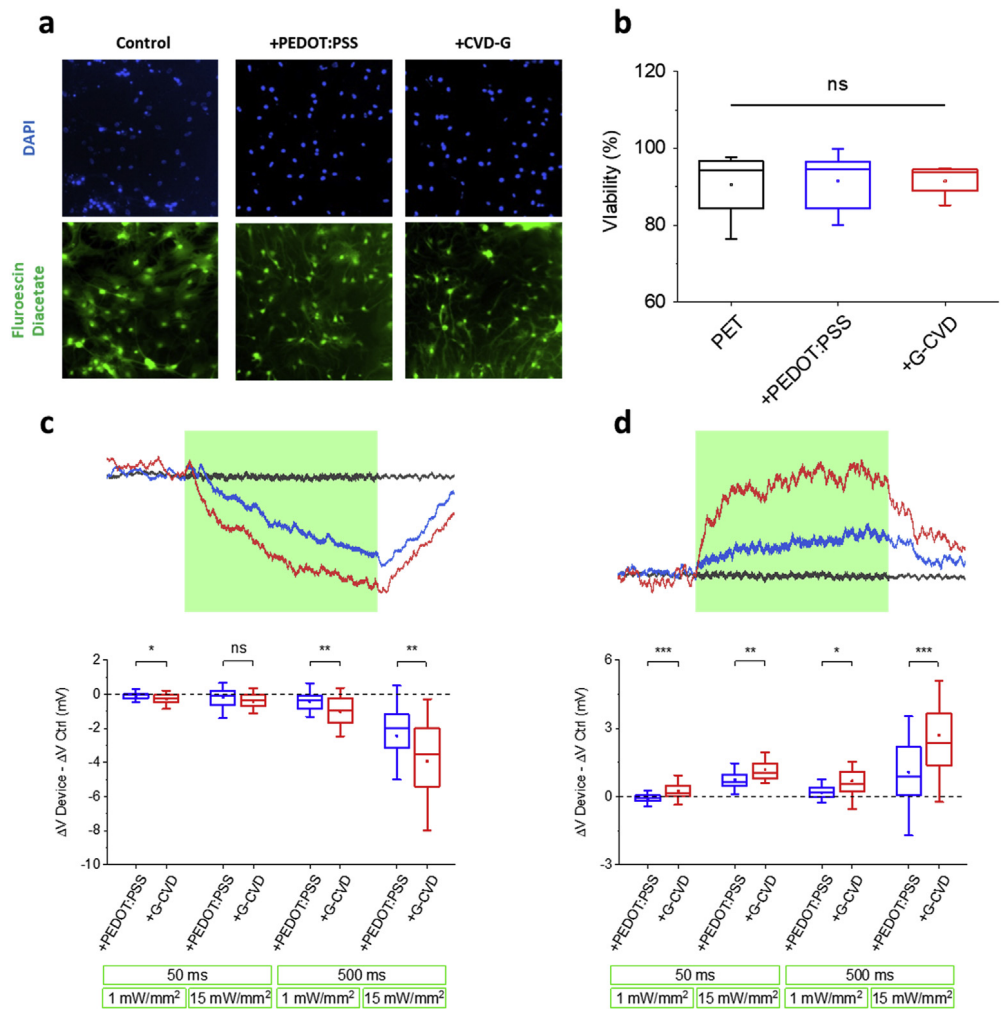

**Fig. 2.** Interaction of primary hippocampal neurons with G-based polymeric devices.
**a**) Viability of primary hippocampal neurons grown onto the various devices. Neurons grown for 14 DIV in contact with PET control, a P3HT/PEDOT:PSS/PET device (+PEDOT:PSS) or a P3HT/G-CVD/PET (+G-CVD) were stained with the DAPI nuclear stain and the live stain Fluorescein Diacetate. **b**) The resulting percent viability was measured as the ratio between the number of Fluorescein Diacetate-positive cells and the total number of cells stained with DAPI. One-way ANOVA/Kruskal-Wallis tests (n = 6, 6, 5 for PET, +PEDOT:PSS and +G-CVD respectively). **c,d**) Light-induced membrane voltage modulation measured in current-clamp configuration with no-current injection (**c**, I = 0) or by setting the holding potential at −70 mV (**d**, $I_{Holding}$ −70 mV). *Upper panels*: Representative traces of primary hippocampal neurons in contact with control PET (black traces), +PEDOT:PSS (blue traces) or + G-CVD (red traces) devices in response to light stimulation at 15 mW/mm$^2$ for 500 ms (green shaded area). *Lower panels:* membrane voltage hyperpolarization (**c**) or depolarization (**d**) in response to various light-stimulation protocols as indicated. Unpaired Student's *t*-test (D'Agostino/Pearson's normal distribution) or Mann-Whitney's *U* test (D'Agostino/Pearson's non-Gaussian distribution) between + PEDOT:PSS and +G-CVD within the same experimental condition (**c**) N = 35, 27, 41, 30, 38, 27, 41, 33; **d**) N = 30, 24, 30, 25, 29, 24, 28, 26 following the order in the plots from left to right). (A colour version of this figure can be viewed online).

neuronal activity more efficiently, particularly when neurons are uniformly polarized at −70 mV.

### 2.4. The enhanced photostimulation of primary neurons with the G-based device shows current and voltage dependence

Taken into account the bimodal response of membrane voltage modulation to light stimuli, we asked whether the same feature was preserved over a wide range of injected currents and membrane voltages. To this aim, primary hippocampal neurons were subjected to patch-clamp measurements in the CC configuration ranging from −300 pA to +300 pA, and in the voltage-clamp configuration (VC) ranging from −110 mV to +10 mV. The current injection or the membrane voltage was maintained all along the recorded sweeps, before, during and after light-stimuli that were applied using the same protocols described above (15 mW/mm$^2$ for 500 ms; Fig. 3).

In the CC configuration, we observed an increasing depolarization towards negative clamped currents and, on the contrary, increasing hyperpolarization towards positive clamped currents (Fig. 3a; blue line and red line for + PEDOT:PSS and +G-CVD respectively). Such a current-dependency of the light-evoked membrane voltage response could be described for neurons in contact with both the +PEDOT:PSS and the +G-CVD devices (Fig. 3c, left panel), and was significantly higher with G-CVD at −300 pA. On the contrary, no current injection-dependency was described for neurons cultured in contact with PET control devices (Fig. 3a,c; black). With less intense and shorter light stimuli, the +G-CVD interface outperformed the +PEDOT:PSS one: while small and non-significant responses were observed to 1 mW/mm$^2$ for 50 ms for both interfaces, 15 mW/mm$^2$ for 50 ms were sufficient for achieving a significant voltage modulation for the +G-CVD interface (Suppl. Figure 2).

The same dependency was maintained when switching the measurement from CC to VC configuration. As expected, light stimuli induced an inward current at hyperpolarizing voltages, that became outward during depolarization (Fig. 3b). When the I/V curves of the light-triggered current were plotted as a function of the clamped voltage (Fig. 3c; right panel), no significant differences were found between + PEDOT:PSS and +G-CVD interfaces, despite a tendency towards a higher light-sensitivity of +G-CVD interfaces at very negative voltages. On the contrary, when we calculated the slope of I/V curves per each cell, we found a significantly stronger improvement of current modulation from neurons in contact with the +G-CVD devices with respect to the +PEDOT:PSS benchmark (Fig. 3d). As described for CC measurements, we repeated experiments at various intensities and/or durations of the light stimuli also in the case of VC configuration (Suppl. Figure 1). Under this configuration, we found that G-CVD improved the light-triggered currents with respect to the +PEDOT:PSS device. The enhancement was significant with stimuli at 1 mW/mm$^2$ for 500 ms or at 15 mW/mm$^2$ for 50 ms, while only a trend was present at 1 mW/mm$^2$ for 50 ms.

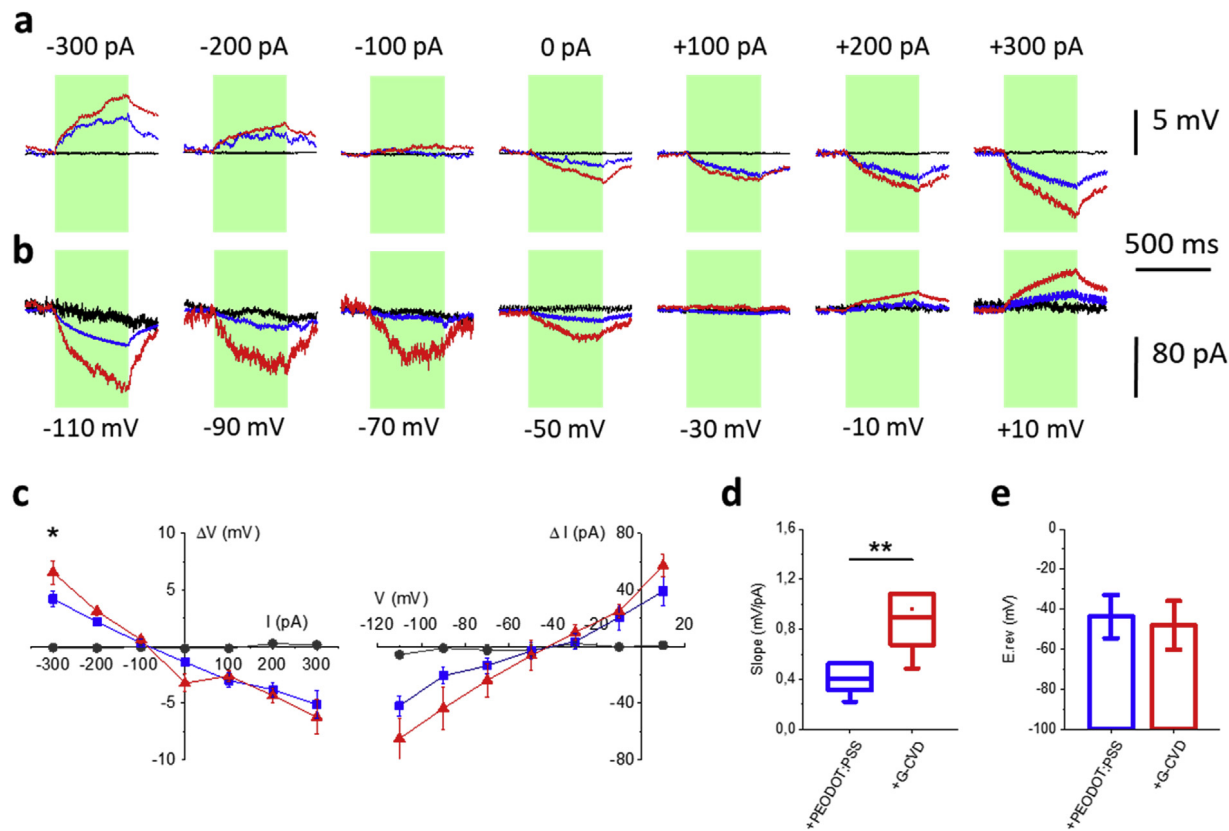

**Fig. 3.** Current- and voltage-dependence of light-induced modulation of membrane potential in G-based polymeric devices.
**a,b)** Representative traces of light-induced voltage (**a**) and current (**b**) modulation in primary hippocampal neurons grown on PET control (black), +PEDOT:PSS (blue) and +G-CVD (red) devices and subjected to light stimulation at 15 mW/mm$^2$ for 500 ms (green shaded area). **c**) I/V curves showing the light-induced voltage modulation (15 mW/mm$^2$ for 500 ms) measured in the current-clamp (CC) configuration from −300 to +300 pA (*left*), and the current modulation measured in the voltage-clamp (VC) configuration from −110 to +10 mV (*right*). Two-way ANOVA/Tukey's tests (CC–I/V, N = 7, 15, 13 for PET, +PEDOT:PSS and +G-CVD respectively; VC-IV N = 6, 9, 8 for PET, +PEDOT:PSS and +G-CVD respectively). **d**) Slope of single-cell current modulation extrapolated from linear fitting of V/C–I/V curves. Mann-Whitney's *U* test (Rout test for outliers, Q = 1%) (N = 7, 8 for + PEDOT:PSS and +G-CVD respectively). **e**) Reversal membrane voltage of the light-triggered current extrapolated from V/C–I/V curves. Unpaired Student's *t*-test (N = 9, 8 for + PEDOT:PSS and +G-CVD respectively). (A colour version of this figure can be viewed online).

To better elucidate whether membrane ionic conductances were involved in phototransduction, we calculated, per each cell, the reversal voltage ($E_{rev}$) of the light-triggered current (Fig. 3e). We found that the reversal potentials of the light-evoked effects of the two devices were closely similar. This suggests that the same cellular mechanism underlies the generation of voltage-dependent currents under light-stimulation, and that the enhanced effect of G-CVD is attributable to its improved capability to separate the charges generated in the P3HT photosensitive layer.

### 2.5. G-CVD increases the efficiency of light-induced firing modulation in primary hippocampal neurons

We previously reported the ability of PEDOT:PSS/P3HT-based neuronal interfaces to decrease spontaneous neuronal firing [15] as a consequence of the hyperpolarization induced by light-stimuli. In the context of the enhanced membrane voltage modulation of the +G-CVD device, a more powerful inhibition of neuronal firing could be expected. To assess this aspect, we cultured primary hippocampal neurons on either + PEDOT:PSS or + G-CVD devices at high density, in order to increase network activity and basal neuronal firing, and recorded action potentials (APs) in the CC configuration at I = 0, using the previously described illumination protocols. Peristimulus time histograms (PSTHs, Bin size 25 ms) were then computed to describe the AP firing under illumination.

The analysis revealed a pronounced inhibition of neuronal activity during light stimulation (Fig. 4a and b). Such a modulation of AP firing in response to the light-evoked hyperpolarization was not homogeneously distributed in the neuronal populations, with some neurons that were insensitive to illumination. Recorded neurons were therefore sorted into responding cells, showing light-induced decrease in firing, and non-responding cells, whose firing was unaffected by illumination. While the ratio between responding and non-responding neurons was similar between the two hybrid devices (Fig. 4c), PSTH analysis performed during the illumination period with a bin size of 500 ms (0−0.5 s from light-onset) revealed a significant more intense firing inhibition with the G-CVD-based device with respect to the +PEDOT:PSS device (Fig. 4d).

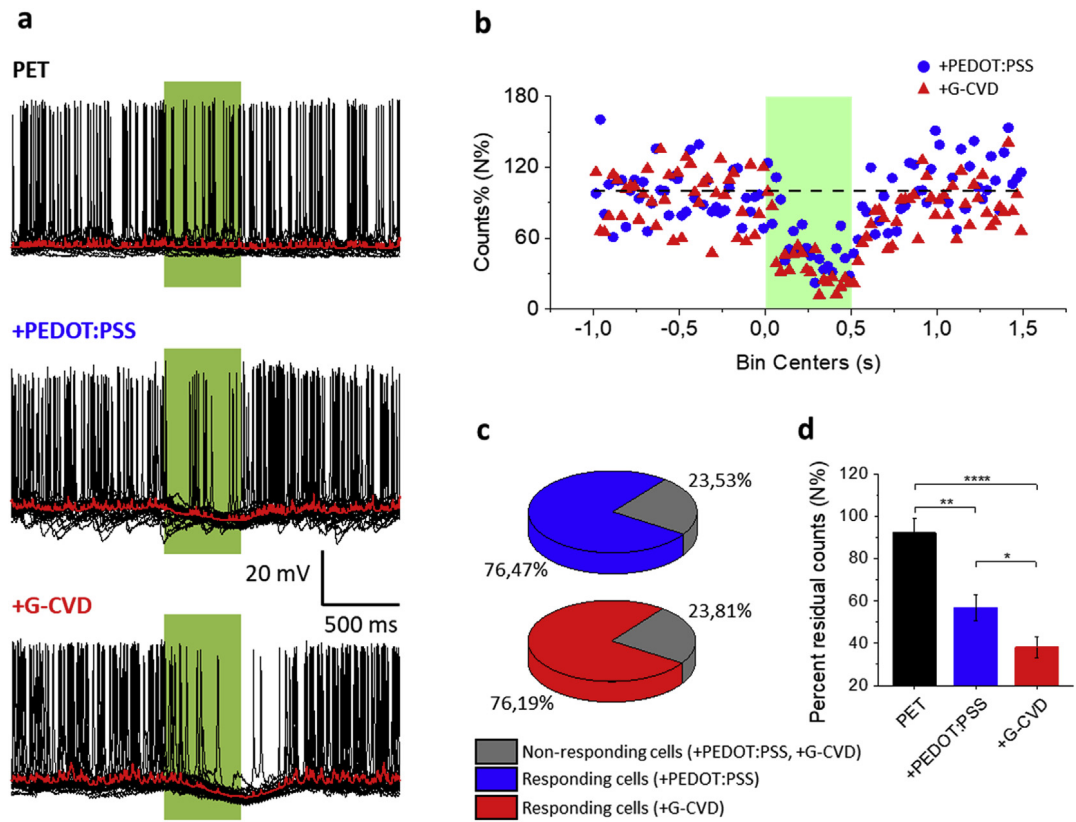

**Fig. 4.** Enhanced light-dependent firing modulation in G-based polymeric devices.
**a)** Representative traces of action potentials (black line) from hippocampal neurons grown on PET (top), +PEDOT:PSS (middle) and +G-CVD (bottom) devices under light stimulation at 15 mW/mm$^2$ for 500 ms (green box). Red lines represent average traces, showing the light-induced hyperpolarization. **b)** Peristimulus Time Histogram (PSTH) analysis (bin size = 25 ms) of action potential firing normalized to the pre-pulse firing activity (from −1.5 to 0 s). Green-shaded area correspond to light-stimuli at 15 mW/mm$^2$ for 500 ms; dashed line corresponds to 100%. N = 13 and 16 light-responding cells for + PEDOT:PSS and +G-CVD, respectively. **c)** Pie-charts of neurons with no-response to light stimulation (grey) *versus* neurons responding to light-stimuli with a decrease in firing rate (colour) on + PEDOT:PSS devices (blue; 13 responding cells *versus* 4 non-responding cells) or + G-CVD devices (red; 16 responding cells *versus* 5 non-responding cells). **d)** PSTH analysis (bin size = 500 ms; means ± SEM) of action potential firing during the 500 ms light stimulus. Two-way ANOVA/Tukey's tests (n = 5, 13, 16 for PET, +PEDOT:PSS and +G-CVD, respectively). (A colour version of this figure can be viewed online).

### 2.6. G-based devices efficiently recover light-sensitivity in blind retina explants

In order to measure the degree of neural activity modulation exerted by devices, we recorded extracellular action potentials elicited by a flash of light onto explanted dystrophic retina from 12 to 14 months-old dystrophic RCS rats, an experimental model of *Retinitis Pigmentosa* in which photoreceptor degeneration is due to a mutation in the *Mertk* gene [17,20]. For the recordings, the explanted retina was placed with the retinal ganglion cells (RGCs) in contact with multielectrode array (MEA) electrodes (epiretinal recording) with the devices layered on the external retina in place of the degenerated photoreceptors. In this subretinal implant configuration, the P3HT surface was in contact with the remnant of the outer plexiform layer and the inner nuclear layer (INL) represented by bipolar cells that are not directly affected by the degenerative process. The subretinal configuration of the device, with light passing through the inverted microscope and reaching the RGC layer first, recapitulates the physiological pathway of visual stimulation that reaches the light sensitive outer retina after crossing the inner retina layers. While no significant effect was present in the absence of any device or with PET alone substrates, a significant firing modulation was observed in the presence of P3HT-based prototypes in response to light stimulation, as shown by PSTH analysis (Fig. 5a). Interestingly, +G-CVD triggered a robust light-dependent modulation of the overall firing rate of RGCs that, at the highest power (500 ms, 540 nm at 41 mW/mm$^2$), was significantly more intense than that observed in the presence of +PEDOT:PSS devices (Fig. 5b).

A better understanding of the dynamics of the effect mediated by the P3HT-based prototypes was obtained by separately considering light-dependent decreases and increases in the firing rate of RGCs, reminiscent of the physiological OFF and ON responses in the retina. Thus, we defined the recorded patterns as OFF-like responses when RGCs were silenced by light stimulation and ON-like responses when RGC firing was stimulated during the light pulse. The two types of light-modulation are represented by the raster plots of Fig. 5c and d for OFF-like and ON-like type of light responses, respectively. The firing modulation was quantified for the +PEDOT:PSS and +G-CVD devices at both 16 and 41 mW/mm$^2$ light power densities during and after the illumination pulse and compared with the basal dark firing rate. OFF-like responses were independent of the type of device employed at both light intensities, and their rebound firing occurring after light offset was significantly reduced with +G-CVD devices at both power densities (Fig. 5c). On the other hand, ON-like responses were strongly and significantly potentiated with +G-CVD devices, as compared with the +PEDOT:PSS devices. Indeed, ON-like RCSs onto + G-CVD devices exhibited a more than two-fold increase in firing during illumination, in the absence of significant effects in the firing activity after light offset (Fig. 5d).

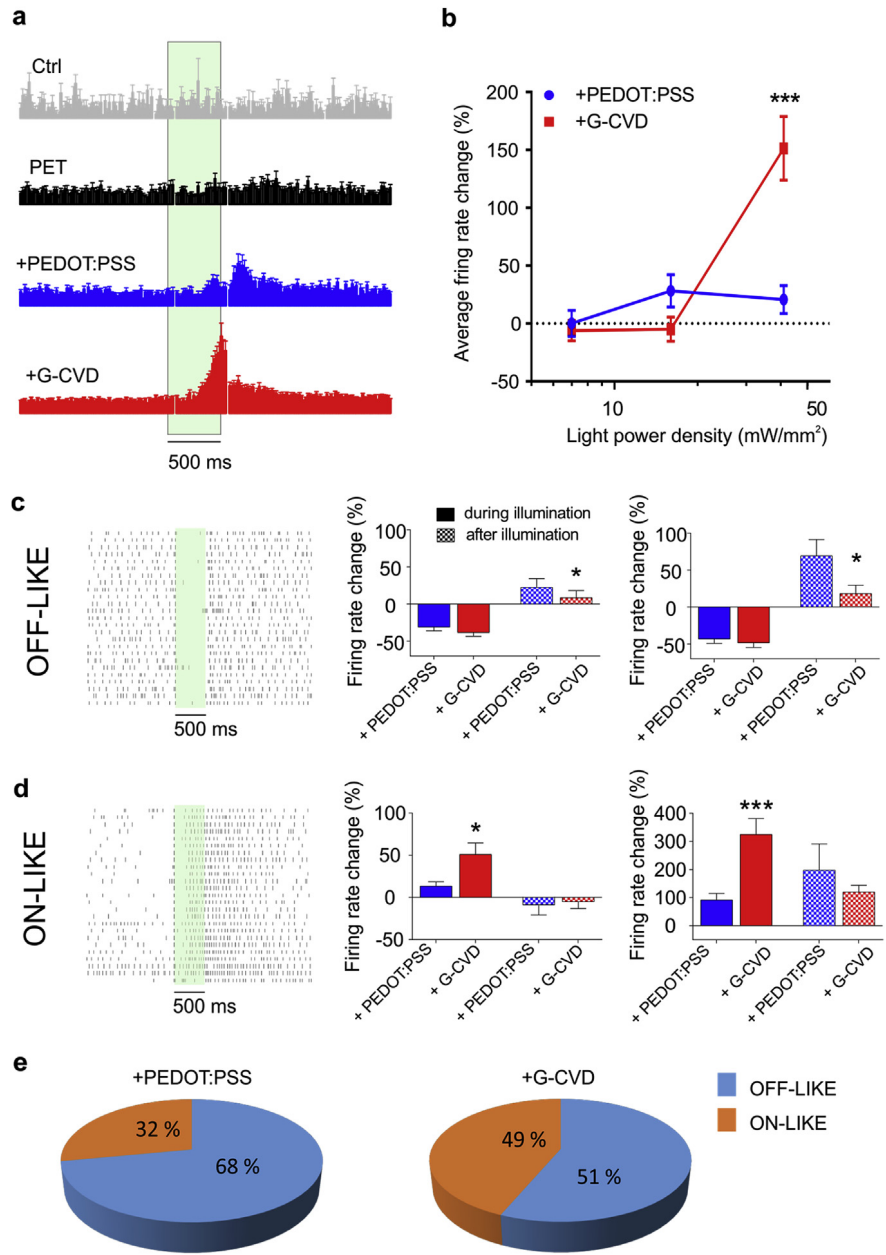

**Fig. 5.** Light triggered ON/OFF-like retinal ganglion cell responses from blind RCS retinas.
**a**) Peristimulus time histograms (PSTHs) (bin size = 20 ms; means ± SEM) of the average RGCs activity epiretinally recorded on MEAs in the presence of PET, P3HT/PEDOT:PSS/PET (+PEDOT:PSS) or P3HT/G-CVD/PET (+G-CVD) positioned subretinally. RGC firing changes in response to light stimuli (green shadowed area; 500 ms at 41 mW/mm$^2$) were normalized to the pre-pulse firing activity (from 0 to 1.5 s). **b**) The average firing rate during the 500 ms of green light pulse is reported as a function of the light power density (more than 20 RGCs for each group from 5 animals, ***p < 0.001 two-way ANOVA/Sidak's tests). **c**) Representative raster plot of an OFF-LIKE light response recorded from a blind RCS retinal explant in subretinal configuration and stimulated by 500 ms (green light at 41 mW/mm$^2$) (*left*). The quantification of the mean firing modulation during and after light stimulation is shown for illumination intensities of 16 mW/mm$^2$ (*middle*) and 41 mW/mm$^2$ (*right*). **d**) The same representation of the firing rate of panel c) for ON-LIKE light responses (n = 5 animals per experimental group, *p < 0.05 ***p < 0.001 Mann Witney's *U* test). **e**) Pie-charts of ON- and OFF-like responses from RGC neurons evoked by light stimulation with subretinal + PEDOT:PSS or + G-CVD devices. (A colour version of this figure can be viewed online.)

When the frequency of occurrence of ON-like and OFF-like responses of the recorded RGC population was analysed with +PEDOT:PSS and G-CVD devices, we found a significant increase of the frequency of ON-like firing, that was accompanied by a corresponding decrease in the OFF-like responses to light with the +G-CVD device. This feature can be particularly interesting in re-establishing a correct balance of ON and OFF responses in animal models of *Retinitis Pigmentosa*, in which the progression of retinal degeneration is associated with inner retina rewiring and decreased ON/OFF RGC ratio, as it is the case of the RCS rat used in this study [21].

## 3. Discussion and conclusions

An increasing interest is growing in the scientific community for novel technologies able to modulate cellular activity using light as a trigger. A major issue is to provide more efficient and spatially resolved technologies, without the drawback of genetic manipulation. In this scenario, organic neuronal interfaces based on conjugated polymers have already proved to be a viable tool for the modulation of neuronal activity *in vitro* [14,15], and for the recovery of visual properties in rodent models of retinal dystrophy [16].

We report here a novel strategy for the improvement of conjugated polymer-based light-sensitive devices by the substitution of the PEDOT:PSS conductive layer with highly conductive G-CVD. Thanks to the more favourable work function of G-CVD for charge extraction with respect to PEDOT:PSS, this architecture allows a more efficient separation of charges generated under illumination in the neighbouring semiconductive P3HT, thereby enhancing neuronal photostimulation. Such a gain of function exhibited by the G-based device allows a more effective modulation of neuronal activity *in vitro*, as described by the analysis of membrane voltage changes and AP firing modulation, as well as *ex vivo* in explants of light-insensitive, degenerate retinas.

Graphene revealed to be effective in improving the efficiency of several devices such as neural interfaces [22] or photovoltaic solar cells [23], and the enhanced work function and consequent neuronal photostimulation reported here stands out another fruitful application of Graphene in the field of neuronal stimulating devices.

Interestingly, the light-induced modulation of neuronal activity appeared to be related to the initial state of membrane voltage: the more hyperpolarized is a neuron, the more pronounced will be the depolarization triggered by illumination and, conversely, the more depolarized is a neuron, the more it will hyperpolarize under illumination. An analogous phenomenon was also found varying the membrane voltage of HEK-293 cells during light-stimulation protocols [13], that was described as a concerted modulation of cellular capacitance, membrane resistance and reversal potential. This feature could bring about an effect of conjugated polymer-based devices, depending on the specific excitation/inhibition balance of the neuronal network subjected to illumination, which has to be taken into account for possible *in vivo* applications, such as retinal implantable devices, where a different modulatory effect could be achieved as a function of the targeted retinal pathway. Such a prosthesis is meant to be implanted *in vivo* in the subretinal configuration, with the photoactive P3HT layer in contact with bipolar cells. In the retina, both photoreceptors and bipolar cells work exclusively by oscillations of the membrane potential, not being able to generate action potentials. Depolarizations and hyperpolarizations occur in response to light in on- and off-bipolar cells, respectively, and mediate activation or silencing of the respective on- and off-RGCs. In this respect, the under-threshold modulations of the membrane potential of bipolar cells by the device can be physiologically relevant for generating light-mediated signals in the degenerate retina devoid of photoreceptors.

*Ex vivo* measurements with explanted samples of blind retina demonstrated that graphene-based stimulation, with respect to PEDOT:PSS, was overall more effective in eliciting light-dependent responses. Furthermore, the results obtained by the new prosthetic prototype were also better resembling an ideal physiological pattern of light response: the ON/OFF-like responses balance, altered by degeneration-induced inner retina rewiring, was re-established by the graphene-based device, differently from the previous device with PEDOT:PSS that was promoting an OFF-like response predominance. This feature is very promising, suggesting that the introduction of graphene could be useful to elicit specific ON or OFF responses, thereby paving the way to further improvements of the device architecture for the design of more efficient retinal prosthetics.

## 4. Methods

### 4.1. Photoactive polymeric devices fabrication

Commercial polyethylene terephthalate (PET) was used as passive substrate for all devices. On top of PET, a conductive layer composed by PEDOT:PSS or G-CVD (Graphenea) was deposited. A water dispersion of PEDOT:PSS (Clevios PH1000; Heraeus) was prepared by adding the following additives: the cosolvent dimethylsulfoxide (9% in volume, purchased from Sigma-Aldrich) to increase the overall electrical conductivity; the crosslinker 3-glycidoxypropyltrimethoxysilane (0.9% in volume; Sigma-Aldrich) to enhance the adhesion of the PEDOT:PSS layer to the substrate and avoid delamination; the surfactant Zonyl FS-300 (0.18% in volume, Sigma-Aldrich) to promote dispersion wettability. PEDOT:PSS dispersion was then sonicated in an ultrasonic bath for 20 min, cooled at room temperature and deposited by spin-coating in two identical steps (rotation speed 2000 rpm, duration 60 s). After deposition, the substrates underwent a thermal annealing process in air (120 °C, 10 min). Monolayers of graphene were synthetized by Chemical Vapour Deposition on a silicon substrate and then transferred to PET (Graphenea) for the preparation of G-based devices. A chlorobenzene solution of P3HT with a regio-regularity of 99.5% (15 000–45 000 molecular weight, Sigma-Aldrich, 30 g/L) was stirred overnight at 70 °C and then deposited on top of the conductive layer by a two-steps spin coating process (800 rpm, 5 s; 1600 rpm, 120 s). A thermal annealing (120 °C, 20 min) completed the fabrication of the devices that were successively cut with a surface of 1.2 $cm^2$ for cyclic voltammetry, patch-clamp measurements or 0.8 $cm^2$ for surface potential analysis, and approximately 0.5 × 0.5 $mm^2$ for MEA measurement. All samples dedicated for cell culture preparation were stuck on glass support with Sylgard 184 silicon elastomer, and sterilized at 180 °C for 2 h.

### 4.2. Electrochemical characterization

Cyclic voltammograms were recorded using Patchmaster V2.73 (HEKA Elektroniks) with a two-electrode setup with copper-coated glass slides on which the prototypes were glued and contacted with silver paste to the P3HT layer. A PTFE tape with a hole of approximately 1 $mm^2$ was employed to expose to the electrolyte the P3HT, acting as the working electrode. All measurements were referenced to an Ag/AgCl electrode. The measurements were carried in phosphate buffer solution at a scan rate of 100 mV/s. Analysis was performed with FitMaster v2x90.1 software, Prism 6.07 (GraphPad) and OriginPro 9 (OriginLab).

### 4.3. Primary neuron preparation

Primary cultures of hippocampal neurons were prepared from embryonic day 18 rat embryos (Charles River). Rats were sacrificed by $CO_2$ inhalation, and embryos removed immediately by cesarean section. Photoactive planar devices (+PEDOT:PSS, +G-CVD, and PET as sham device) were pre-treated with Poly-L-lysine (0.1 mg/ml in borate buffer) prior to cell-plating. Briefly, hippocampi were dissociated by a 30-min incubation with 0.25% trypsin at 37 °C and cells were plated on polymeric devices in Neurobasal supplemented with 2 mM L-glutamine, 2% B27, 100 μg/ml penicillin and 100 mg/ml streptomycin and with 10% horse serum (Life Technologies) in the first 4 h of plating. Low-density cultures were prepared plating 40'000 to 80'000 neurons per device, while high-density cultures were prepared plating 80000 to 160000 neurons per device. All animal manipulations and procedures were performed in accordance with the guidelines established by the European Community Council (Directive 2010/63/EU of March 4th, 2014) and were approved by the Institutional Ethics Committee and by the Italian Ministry of Health.

### 4.4. Surface potential analysis

Photocurrent measurements were performed in voltage-clamp mode at room temperature in recording extracellular solution with patch pipettes (4–6 MV) filled with the same solution. Responses were amplified, digitized at 10 or 20 kHz and stored with Patchmaster V2.73 (HEKA Elektroniks). FitMaster v2x90.1 were employed for data analysis, together with Prism 6.07 (GraphPad) and OriginPro 9 (OriginLab).

### 4.5. Electrophysiology

Whole-cell patch-clamp recordings of hippocampal neurons (between 14 and 18 day *in vitro*, DIV) were performed at room temperature using borosilicate patch pipettes (3.5–5.0 MΩ) and under GΩ patch seal. The extracellular solution contained (in mM): 135 NaCl, 5.4 KCl, 1 $MgCl_2$, 1.8 $CaCl_2$, 5 HEPES, 10 glucose adjusted to pH 7.4 with NaOH. The intracellular solution contained (in mM): 126 K-Gluconate, 4 NaCl, 1 $MgSO_4$, 0.02 $CaCl_2$, 0.1 EGTA, 10 Glucose, 5 Hepes, 3 ATP-$Na_2$, and 0.1 GTP-Na. Recordings were carried out using an EPC10 (HEKA Elektroniks) amplifier. Measurement of membrane voltage modulation under light stimulation were performed in current-clamp configuration with i) no current injection (I = 0 pA), ii) clamping the current necessary to maintain neurons at −70 mV ($I_{H -70 mV}$), iii) clamping currents ranging from −300 to + 300 pA, and iv) in voltage-clamp configuration at voltages from −110 to +10 mV. Responses were amplified, digitized at 10 or 20 kHz and stored with Patchmaster V2.73 (HEKA Elektroniks). FitMaster v2x90.1 were employed for data analysis, together with Prism 6.07 (GraphPad) and OriginPro 9 (OriginLab). Photostimulation of neurons was carried out on a Nikon FN1 upright microscope (Nikon Instruments) by using a combination of 510 nm and 550 nm wavelengths of Spectra X LED system (Lumencor) to match the P3HT absorption spectrum. The light source had a final power of 1 or 15 mW/$mm^2$, with stimulations of 50 or 500 ms in duration.

The slope of current modulation was extracted per each cell through linear fitting of VC-IV curves, as calculated with OriginPro 9. Reversal voltage ($E_{rev}$) was calculated per each cell as the intercept of the VC-IV linear fitted curve on the Y-axis at 0 pA.

### 4.6. Explant procedure for MEA experiments

Following 30 min of dark adaptation, animals were euthanized by means of high-dose carbon dioxide ($CO_2$) inhalation. All the procedures were carried out in dim red light since rats are known to lack photoreceptors responding to this wavelength. Both eyeballs were surgically enucleated and quickly placed in oxygenated (95% $O_2$, 5% $CO_2$) AMES Medium solution (Sigma-Aldrich), minimizing the time in which retinal tissue was left deoxygenated. Dissection of the eyeball was performed first on a dry surface and then in liquid. The procedure is performed as follows: i) a first hole through the eyeball at the level of the *ora serrata* is obtained by means of the tip of a scalpel blade, ii) dissecting scissors are passed through the hole and used to cut and dissect the cornea from the sclera, choroid and retina, iii) lens and vitreous body are removed, iv) residual tissues are cut in half in order to facilitate the following steps and then moved into a separate AMES solution Petri dish in order to continue the dissection under the microscope. In liquid, the two hemiretinas are gently dissected from the sclera by means of dissecting forceps and from the choroid, avoiding pointing or grasping, in order to minimize the stress on the tissue. Because the manipulation may initially alter electrophysiological properties of the tissue, recordings are initiated after about 10 min of rest in oxygenated AMES solution. Each hemiretina is divided into smaller pieces prior to the settling onto the MEAs.

### 4.7. MEA recordings

Recording of extracellular activity and action potentials was carried out on 60-electrodes planar MEA devices. Electrodes of 30 μm in diameter are disposed in a 8x8 matrix, with an inter-electrodes distance of 200 μm; one of the electrodes works as an internal reference (iR) (MEA 200/30iR-ITO-gr, Multi Channel Systems GmbH). The devices were also provided with a 1.5 ml capacity glass ring/culture chamber that allowed perfusing the oxygenated solution. Recording electrodes are made of nanostructured titanium nitride (TiN) featuring impedance less than 100 kΩ (specific for 30 μm Ø electrode). Traces and contacts are made of transparent Indium tin oxide (ITO), allowing for a clear vision of the sample under the microscope. Electrical passivation is provided by silicon nitride. Pieces of retina to be tested were cut out from the previously isolated hemiretinas in AMES' Medium, with arbitrary dimensions matching as closely as possible the electrodes surface on the MEA (~1 $mm^2$).

Experimental samples were placed onto the MEA with the RGCs layer facing the electrodes surface (face down on the plate). The different devices were place directly on top the tissue, with the P3HT layer facing it. Eventually, a plastic net and a metal anchor were further placed on top of the tissue + device to favour a more uniform adhesion. The MEA chamber was finally filled with AMES Medium and transferred to the recording set-up. To keep tissue oxygenated, the MEA chamber was continuously perfused by a peristaltic pump (ISMATEC) with oxygenated (95% $O_2$ 5% $CO_2$) AMES Medium. Electrical activity was recorded using the MEA1060-INV BC by Multichannel Systems (MCS GmbH, Reutlingen, Germany), with a total amplification of 1200 (two stage amplification). MC Rack software (MCS GmbH) was used for the recording, detection and sorting of the data. In order to distinguish action potentials from background noise (~15 μV, peak-to-peak), we arbitrarily chose a threshold of detection equal to 4.5 times the SD of the signal (automatically calculated by the acquisition system for each electrode). The firing activity of retinal ganglion cells has been recorded for 4 s across each light pulse and the intervals of interest were the 500 ms during illumination and the following 500 ms starting at light offset. After each light pulse, the firing increased thanks to P3HT-based prototypes, and returned back to a baseline rate in less than a second. The illumination pulse was set at 0.25 Hz and lasted around 1.5 min (25 sweeps). An integrated device allowed also to keep the temperature of the perfused AMES' Medium and of the MEA plate at a desired value of 37 °C. The MCS system was optically coupled with light stimulation by means of the inverted microscope Nikon Eclipse Ti, using a VSD IR filter and a 20x objective, producing an illumination spot on the sample of 0.93 $mm^2$. The microscope was coupled with a CCD Hamamatsu Orca D2 camera for image capturing. The entire setup was placed on a vibration isolation table and shielded from electromagnetic radiation by a Faraday cage. Light stimulation was provided by the Lumencor Spectra X Light engine, composed of 6 independent solid-state LED that can be individually activated. The excitation beam is coupled to one of the microscope's ports. Light intensity was measured using a power meter ThorLabs PM100D and converted into power density given the area of the illumination spot.

MC-Rack recordings and Spectra X stimulations were synchronized by the Stimulus Generator STG4008 (MCS GmbH). By means of the MC-Stimulus II software it is possible to drive TTL signals (Transistor-Transistor Logic) that synchronize MEA recordings with an external device, in this case the Spectra X light source.

### Declaration of competing interest

All co-authors have agreed to the submission of the final manuscript and have no financial conflict of interest that might be construed to influence the results or interpretation of their manuscript.

### CRediT authorship contribution statement

**Mattia L. DiFrancesco:** Formal analysis, Writing - original draft. **Elisabetta Colombo:** Formal analysis, Writing - original draft. **Ermanno D. Papaleo:** Formal analysis. **José Fernando Maya-Vetencourt:** Validation. **Giovanni Manfredi:** Formal analysis. **Guglielmo Lanzani:** Conceptualization, Writing - review & editing. **Fabio Benfenati:** Conceptualization, Writing - review & editing, Funding acquisition.

### Acknowledgments

We thank D. Mattia Bramini (Istituto Italiano di Tecnologia, Genova, Italy) for help and advice in the neuron viability experiments and R. Ciancio, I. Dall'Orto, A. Mehilli, R. Navone and D. Moruzzo (Istituto Italiano di Tecnologia, Genova, Italy) for technical assistance. This project has received funding from the European Union's Horizon 2020 research and innovation programme under grant agreements No. 696656 (Graphene Flaghship - Core 1) and No. 785219 (Graphene Flaghship - Core 2).

### Appendix A. Supplementary data

Supplementary data to this article can be found online at https://doi.org/10.1016/j.carbon.2020.02.043.